\documentclass{aa}
\usepackage{graphicx}
\usepackage{txfonts}
\usepackage{natbib}
\usepackage{color}
\usepackage[colorlinks, allcolors=blue]{hyperref}

\begin{document}

\title{Evolution Of Downflows In The Transition Region\\ Above A Sunspot Over Short Time-Scales}

\author{C. J. Nelson$^{1}$, S. Krishna Prasad$^{2}$, M. Mathioudakis$^{1}$}

\offprints{c.nelson@qub.ac.uk}
\institute{$^1$Astrophysics Research Centre (ARC), School of Mathematics and Physics, Queen’s University, Belfast, BT7 1NN, NI, UK.\\
$^2$Centre for mathematical Plasma Astrophysics, KU Leuven, Celestijnenlaan 200B, 3001 Leuven, Belgium.}

\date{}

\abstract
{Downflows with potentially super-sonic velocities have been reported to occur in the transition region above many sunspots; however, how these signatures evolve over short time-scales in both spatial and spectral terms is still unknown and requires further research.}
{In this article, we investigate the evolution of downflows detected within spectra sampling the transition region on time-scales of the order minutes and search for clues as to the formation mechanisms of these features in co-temporal imaging data.}
{The high-resolution spectral and imaging data used to identify downflows here were sampled by the Interface Region Imaging Spectrograph on the $20$th and $21$st May $2015$. Additionally, photospheric and coronal imaging data from the Hinode and Solar Dynamics Observatory satellites were studied to provide context about the wider solar atmosphere.}
{Four downflows were identified and analysed through time. The potential super-sonic components of these downflows had widths of around $2$\arcsec\ and were observed to evolve over time-scales of the order minutes. The measured apparent downflow velocities were structured both in time and space, with the highest apparent velocities occurring above a bright region detected in \ion{Si}{IV} $1400$ \AA\ images. Downflows with apparent velocities below the super-sonic threshold assumed here were observed to extend a few arcseconds away from the foot-points suggesting that the potential super-sonic components are linked to larger-scale flows. The electron density and mass flux for these events were found to be within the ranges of $10^{9.6}$-$10^{10.2}$ cm$^{-3}$ and $10^{-6.81}$-$10^{-7.48}$ g cm$^{-2}$ s$^{-1}$, respectively. Finally, each downflow formed at the foot-point of thin `fingers' extending out around $3$-$5$\arcsec\ in \ion{Si}{IV} $1400$ \AA\ data with smaller widths (<$1$\arcsec) than the super-sonic downflow components.}
{Downflows can appear, disappear, and recur within time-scales of less than one hour in sunspots. As the potential super-sonic downflow signatures were detected at the foot-points of both extended fingers in \ion{Si}{IV} $1400$ SJI data and sub-sonic downflows in \ion{Si}{IV} $1394$ \AA\ spectra, it is likely that these events are linked to larger-scale flows within structures such as coronal loops.}

\authorrunning{Nelson et al.}
\titlerunning{Super-sonic downflows in sunspots}

\maketitle

\section{Introduction}
	\label{Introduction}

Sunspots are one of the most widely studied features in the zoo of solar phenomena. Although sunspots themselves can be (relatively) stable over the course of days, many transient events with lifetimes of the order minutes to hours have been observed to occur within them. In the lower solar atmosphere (photosphere and chromosphere), features such as umbral flashes (\citealt{Beckers69}), short dynamic fibrils (\citealt{Rouppe13}), and small-scale umbral brightenings (\citealt{Nelson17}) are evident in a range of spectral diagnostics. Higher in the atmosphere (the transition region and corona), bright umbral dots (\citealt{Tian14bd}), shock wave behaviour (\citealt{Tian14}), and, of most interest here, downflows which would be super-sonic at transition region temperatures (\citealt{Dere82}) have been detected in a large number of sunspots. For a recent review of our understanding of the transition region above sunspots see \citet{Tian18}.

The presence of both apparent sub-sonic and super-sonic downflows above sunspots in spectral lines forming at transition region temperatures was first reported nearly fourty years ago using data from the High Resolution Telescope and Spectrograph (HRTS; see \citealt{Dere82, Nicolas82}). Subsequent research has shown that such downflows can be persistent within sunspots over multiple days (\citealt{Kjeldseth88, Kjeldseth93, Nelson20}), can have apparent velocities of over $200$ km s$^{-1}$ (\citealt{Brekke90, Kleint14}), and can form at the foot-points of coronal loops rooted in sunspot umbrae (\citealt{Chitta16, Nelson20}).  As these events, more recently referred to as `dual flows' (\citealt{Brynildsen01}), can be detected in the majority of sunspots, above both the penumbra and umbra (\citealt{Brynildsen04, Samanta18}), improving our understanding of downflows, with both low and high velocities, is imperative if we are to properly understand sunspots in general.

\begin{figure*}
\includegraphics[width=0.99\textwidth,trim={2.7cm 0 0 0}]{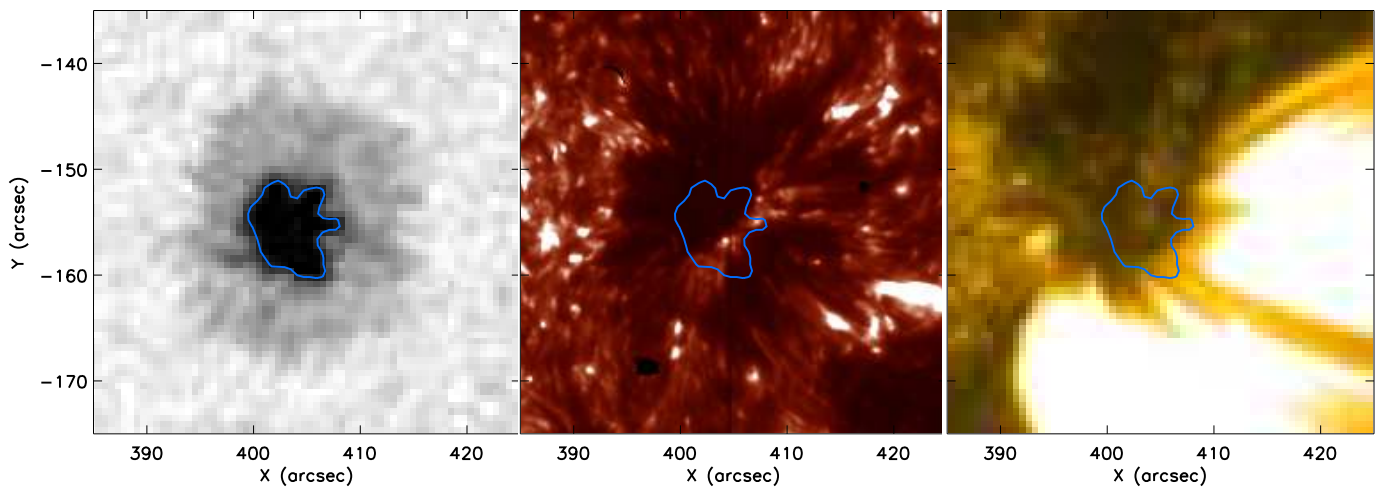}
\caption{The sunspot studied here as observed by the SDO/HMI continuum (left), IRIS $1400$ SJI (middle), and SDO/AIA $171$ \AA\ (right) channels at approximately $13$:$56$:$30$ UT on the $21$st May $2015$. The blue contours on each panel outline the umbra of the sunspot as inferred from the SDO/HMI continuum image.}
\label{Overview_df}
\end{figure*}

The launch of the Interface Region Imaging Spectrograph (IRIS; \citealt{dePontieu14}) has allowed numerous advances to be made in our comprehension of transition region downflows over recent years. Specifically, it is now thought that at least two different types of spectra associated with downflows which would be super-sonic at transition region temperatures appear to be present in IRIS data. One type of spectra manifests as transient red-shifted broadening of spectral lines, sampling from the chromosphere to the transition region, to apparent velocities of over $200$ km s$^{-1}$. The second type of spectra associated with apparent super-sonic downflows is typically observed in the \ion{Si}{IV} and \ion{O}{IV} lines with a clear separation between the line core intensity and a secondary emission peak in the red wing of the line (\citealt{Tian14, Straus15, Nelson20}). These spectra display typical apparent downflow velocities of between $50$-$150$ km s$^{-1}$ and do not always correspond to secondary emission peaks in the lower temperature lines, such as the \ion{Mg}{II} and \ion{C}{II} lines (see, for example, \citealt{Samanta18}). Although these two types of spectra are observationally distinct, it has been suggested that both could be formed as a response to flows forming in the solar atmosphere, perhaps due to the on-set of thermal non-equilibrium (\citealt{Antolin20}). This process could lead to the formation of coronal rain which would then fall along the magnetic field lines into sunspot umbrae (see, for example: \citealt{Kleint14, Ishikawa20}). The first kind of spectra could, therefore, form as a response to bursty coronal rain and the second type could appear due to longer-lived rain events, perhaps caused by aborted condensations (\citealt{Mikic13}). This interpretation is supported by the similarities between both the velocities and densities of these downflows and coronal rain (e.g., \citealt{Antolin12, Antolin15}).

One of the currently unexplored aspects of transition region downflows is how individual regions evolve through time in combined spatial and spectral terms. The reason for this is relatively simple. As IRIS is a slit based instrument, it can either raster or sit-and-stare. A raster scan may sample the two-dimensional structuring of a downflow but if it includes too many steps the temporal evolution of the event on time-scales of the order minutes will remain hidden (as was the case for both \citealt{Samanta18, Nelson20}). Conversely, if the slit is in sit-and-stare mode, then one will be able to infer the spectral evolution of the downflow in exquisite detail but no inferences about the two-dimensional structuring of the event can be made (see \citealt{Straus15, Nelson20}). As examples of both stable (\citealt{Straus15}) and variable (\citealt{Chitta16, Nelson20}) transition region downflows have been observed, further analysis is required to better understand the short-term dynamics of these events.

In this article, we analyse the spatial structuring of four downflow events, with both sub-sonic and super-sonic components (assuming transition region temperatures), over time-scales of the order minutes using two datasets collected by the IRIS satellite. The aim of this work is to investigate how transition region downflows evolve within a sunspot and to better understand what their formation mechanims could be. Our work is set out as follows: In Sect.~\ref{Observations} we detail the datasets studied in this article; In Sect.~\ref{Results} we present our results; In Sect.~\ref{Conclusions} we provide a discussion; Before in Sect.~\ref{Summary} we summarise the key results.

\section{Observations}
	\label{Observations}

Two datasets consisting of $16$-step dense (step-size of $0.35$\arcsec) rasters sampled by the IRIS satellite centred on the lead positive-polarity sunspot within AR $12348$ were used to study downflows. The step cadence was approximately $32$ s (exposure time of $\sim30$ s) giving a total raster cadence of $\sim510$ s. The spatial sampling along the slit was $0.166$\arcsec\ and the spectral sampling was $0.0254$ \AA\ and $0.0127$ \AA\ for the near (\ion{Mg}{II}) and far (\ion{Si}{IV} and \ion{C}{II}) UV windows, respectively. The slit-jaw imager (SJI) sampled the \ion{Si}{IV} $1400$ \AA\ and \ion{Mg}{II} $2796$ \AA\ channels with a cadence of $64$ s and an approximate pixel scale of $0.166$\arcsec. The OBSID for these datasets was: $3800013483$. The first dataset (in chronological order) was collected between $09$:$27$:$43$ UT and $13$:$42$:$34$ UT on the $20$th May $2015$ and consisted of $30$ raster repeats. The second was collected between $11$:$24$:$43$ UT and $13$:$57$:$38$ UT on the $21$st May $2015$ and contained $18$ raster repeats.  Additionally, images co-temporal to the first and final frames of the SJI datasets were downloaded for the Solar Dynamics Observatory's Helioseismic and Magnetic Imager (SDO/HMI; \citealt{Scherrer12}) continuum and Atmospheric Imaging Assembly (SDO/AIA; \citealt{Lemen12}) $171$ \AA\ channels. These data have a post-reduction pixel scale of around $0.6$\arcsec. Finally, high-resolution ($0.054$\arcsec\ pixel scale) imaging data from Hinode's Solar Optical Telescope (SOT; \citealt{Tsuneta08}) co-temporal to the first IRIS dataset were studied to provide further context.

\begin{figure*}
\includegraphics[width=0.99\textwidth]{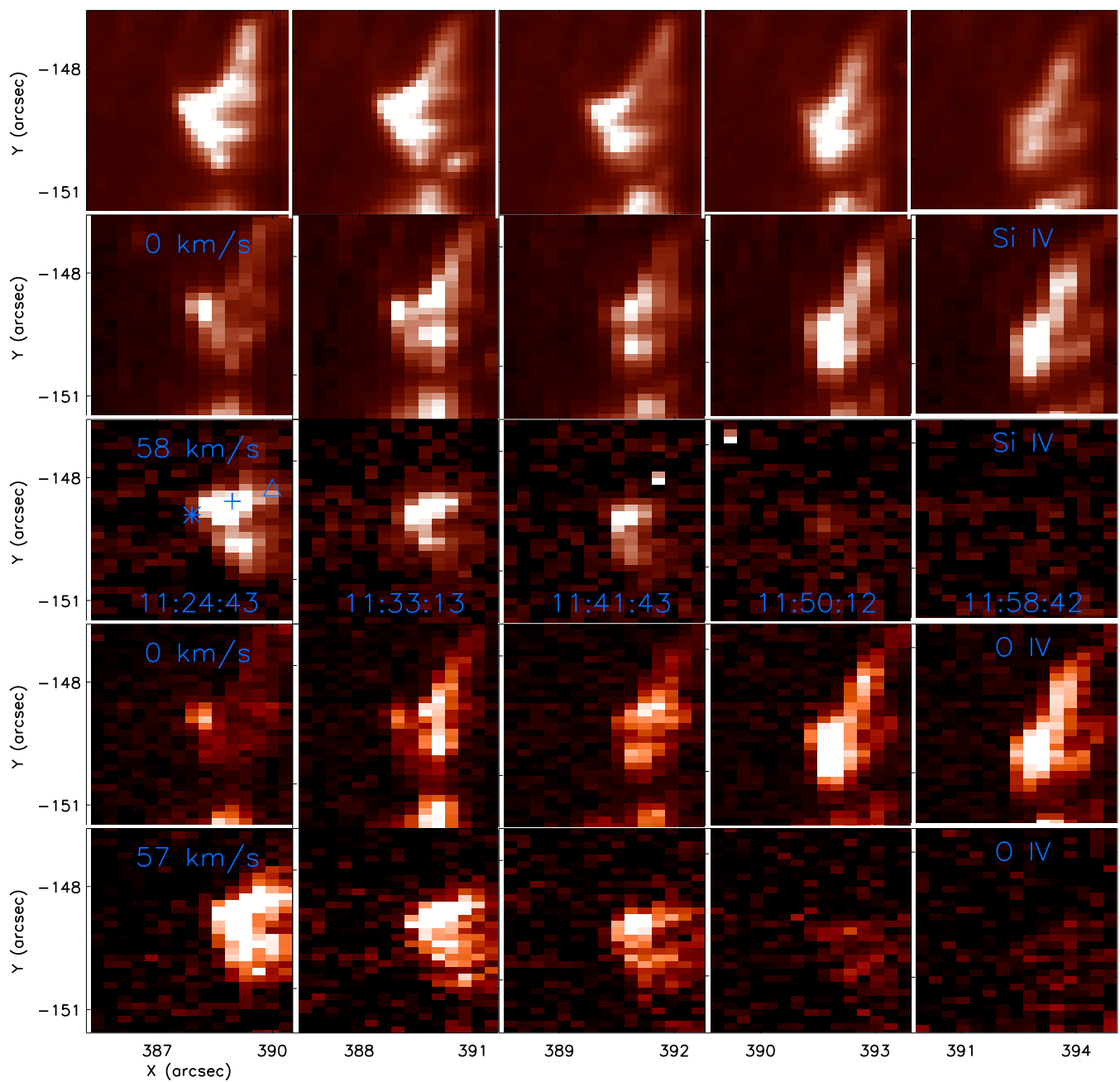}
\caption{(Top row) The evolution of Event (a) through time as sampled by the IRIS $1400$ \AA\ SJI channel. (Second row) The \ion{Si}{IV} $1394$ \AA\ intensity at the line rest wavelength co-spatial and co-temporal to the SJI images. (Third row) Same but for spectral position corresponding to an apparent Doppler velocity of $58$ km s$^{-1}$ in the red wing of the \ion{Si}{IV} $1394$ \AA\ line. (Fourth row) Equivalent plots for the \ion{O}{IV} $1401$ \AA\ line core. (Bottom row) Same as above but for $57$ km s$^{-1}$ into the red wing of the \ion{O}{IV} $1401$ \AA\ line. The downflow can be identified easily as the region of increased brightness in the third and bottom rows. The symbols over-laid on the middle left panel indicates the pixels whose spectra are plotted through time in Fig.~\ref{Spectra1_df}.}
\label{Evolution1_df}
\end{figure*}

In Fig.~\ref{Overview_df} we plot the sunspot analysed here at approximately $13$:$56$:$30$ UT on the $21$st May $2015$ as sampled by the SDO/HMI continuum (left panel), IRIS $1400$ \AA\ SJI filter (middle panel), and SDO/AIA $171$ \AA\ channel (right panel). The blue contours over-laid on each panel outline the sunspot as inferred from the SDO/HMI continuum channel. No lightbridge is evident in the photosphere, however, a bright structure appears to cross the umbra in the transition region (from the south-west to the north-east). In the SDO/AIA $171$ \AA\ channel, fan loops can be detected to the south and east of the sunspot. These fan loops are potentially rooted in the bright transition region structure evident in the IRIS $1400$ \AA\ channel, however, higher resolution observations would be required to confirm this. Within the two datasets studied here, four distinct downflow events were identified and analysed manually using the CRISPEX IDL package (\citealt{Vissers12}). Here, the cut-off for what we denote as `super-sonic' apparent velocities was calculated using the formula $c_\mathrm{s}$=$0.152T^{0.5}$ km s$^{-1}$ (\citealt{Priest84}), with $c_\mathrm{s}\approx48$ km s$^{-1}$ being returned for a temperature of $10^5$ K. As the \ion{Si}{IV} $1394$ \AA\ line has a typical formation temperature of around $T$=$10^{4.8}$ we consider any apparent downflows with velocities over $50$ km s$^{-1}$ to be super-sonic in this article. Of course, whether these events are actually super-sonic depends on various other factors within the plasma.

\section{Results}
	\label{Results}

\subsection{Morphology of transition region downflows}

\subsubsection{Event (a) - disappearance of a transition region downflow}

\begin{figure*}
\includegraphics[width=0.99\textwidth]{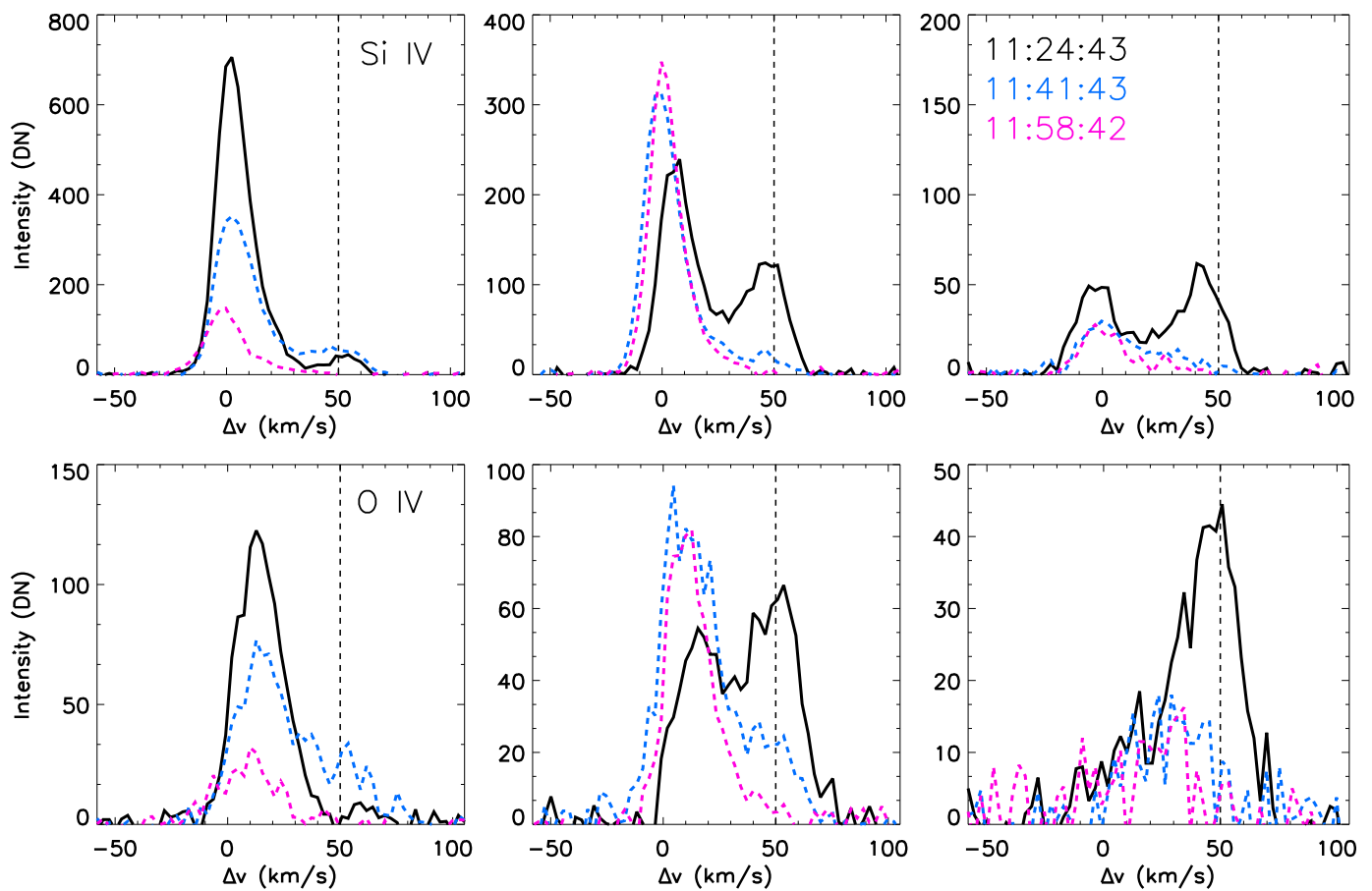}
\caption{The spectral profiles sampled at the locations of the three symbols  on Fig.~\ref{Evolution1_df} for the first, third, and fifth rasters for the \ion{Si}{IV} $1394$ \AA\ (top row) and \ion{O}{IV} $1401$ \AA\ (bottom row) lines. The ordering of the panels is the star, the cross, and then the triangle from left to right, respectively. These plots clearly depict the presence of the distinct secondary emission peak associated with the downflow in the first raster (solid black lines) and its disappearance through time (blue dashed lines then pink dashed lines). The vertical dashed line indicates an apparent downflow velocity of $50$ km s$^{-1}$, used as a threshold for super-sonic downflows in the \ion{Si}{IV} $1394$ \AA\ line in this study, for reference.}
\label{Spectra1_df}
\end{figure*}

The first downflow event analysed here, `Event (a)', was observable close to the umbra-penumbra boundary of the sunspot, to the east of the umbra, in the first three to four rasters of the dataset sampled on the $21$st May $2015$. In Fig.~\ref{Evolution1_df}, we plot the evolution of Event (a) during this time for the \ion{Si}{IV} $1400$ \AA\ SJI filter (top row), the \ion{Si}{IV} $1394$ \AA\ rest wavelength ($0$ km s$^{-1}$; second row), $58$ km s$^{-1}$ into the red wing of the \ion{Si}{IV} $1394$ \AA\ line (middle row), the \ion{O}{IV} $1401$ \AA\ rest wavelength (fourth row), and $57$ km s$^{-1}$ into the red wing of the \ion{O}{IV} $1401$ \AA\ line (bottom row). Time-stamps on the panels in the middle row indicate the start-time of that specific raster. In the initial time-step in the SJI row, a bright structure can be detected with a near-triangular base (with a width of around $2$\arcsec) and two long `fingers' (with widths of <$1$\arcsec) extending up in a curved manner towards the top right corner of each panel. It is immediately evident that the apparent size and intensity of this structure gradually decreased over these $34$ minutes. In the second row of Fig.~\ref{Evolution1_df}, however, where the \ion{Si}{IV} $1394$ \AA\ rest wavelength is depicted, the opposite behaviour is observed. The bright region was initially small and point-like, before it became more elongated (as in the SJI data) with an increase in size and intensity of the structure being measured from raster to raster. 

The downflow at apparent super-sonic velocities present at this location is evident as the bright region in the red wing of the \ion{Si}{IV} $1394$ \AA\ line (middle row) in the first three columns. Interestingly, this region appeared to be spatially split into two distinct `arms' (reminiscent of the flame-like structuring often observed in Ellerman bombs in the H$\alpha$ line wings; see, for example, \citealt{Vissers15, Nelson15}), each forming at the foot-points of the two fingers apparent in the SJI data. These arms were separated by around $0.166$\arcsec-$0.498$\arcsec\ ($1$-$3$ pixels) along their lengths. Downflows with apparent super-sonic velocities were still detected between the arms, however, the magnitude of the secondary emission peak in the red wing was much reduced. The apparent length of these arms was between $1$-$2.5$\arcsec\ and the width of each arm was just over $1$\arcsec. It should be noted that dual flows (not always super-sonic) were detected up to the edge of the raster and, therefore, Event (a) could be larger than measured here. Over the course of the following $34$ minutes, the downflow completely disappeared at this location (see right-hand panel of the middle row of Fig.~\ref{Evolution1_df}).

The behaviour of this downflow in the \ion{O}{IV} $1401$ \AA\ line core (fourth row of Fig.~\ref{Evolution1_df}) was very similar to the behaviour previously described for the \ion{Si}{IV} $1394$ \AA\ line. At the \ion{O}{IV} $1401$ \AA\ rest wavelength, Event (a) was initially small and point-like before it evolved to reveal evidence of its elongated nature, with two fingers appearing to emanate from the triangular foot-point in the final column. The two distinct arms of the super-sonic downflow were evident in the red wing of the \ion{O}{IV} $1401$ \AA\ line at $57$ km s$^{-1}$ in the initial raster (bottom row of Fig.~\ref{Evolution1_df}); however, the lower arm was no longer detectable in the third raster and the entire event had completely faded from view by the time the fourth and fifth rasters were sampled. No secondary emission peak associated with this downflow was apparent in the red wings of either the \ion{Mg}{II} or \ion{C}{II} lines.

\begin{figure*}
\includegraphics[width=0.99\textwidth]{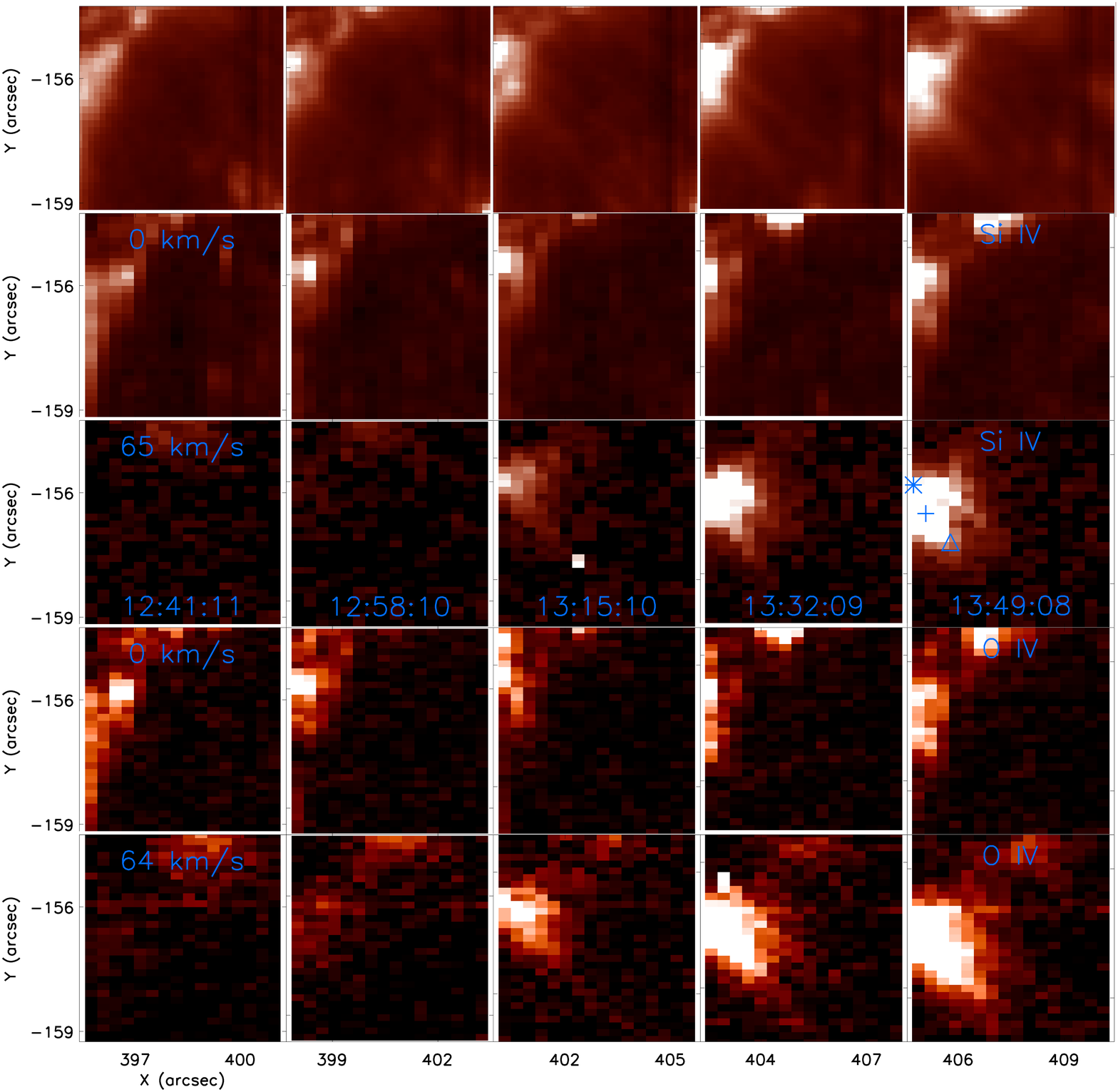}
\caption{Same as Fig.~\ref{Evolution1_df} but for Event (b) and with the red wing images being sampled at $65$ km s$^{-1}$ and $64$ km s$^{-1}$ for the \ion{Si}{IV} $1394$ \AA\ and \ion{O}{IV} $1401$ \AA\ lines, respectively.}
\label{Evolution2_df}
\end{figure*}

In Fig.~\ref{Spectra1_df}, a spectral representation of these dynamics is plotted for the three locations indicated by the symbols on the left panel of the middle row of Fig.~\ref{Evolution1_df}. The vertical dashed line is located at $50$ km s$^{-1}$ for reference. The peak apparent velocity of this downflow in the \ion{Si}{IV} $1394$ \AA\ (top row) and \ion{O}{IV} $1401$ \AA\ (bottom row) lines, approximately $57$ km s$^{-1}$, was measured at the base of the event (left-hand panels; star) in the first raster (black solid line). By the time the fifth raster was recorded $34$ minutes later (pink dashed line), the peak intensity of the line core had reduced by a factor of four and the dual-flow had completely disappeared. No clear reduction in the downflow velocity was detected prior to its disappearance at any of these locations, however higher cadence data would be required to completely rule out whether any deceleration occurred. The increase in intensity in the \ion{Si}{IV} $1394$ \AA\ rest wavelength discussed previously was detected in the middle of the event (top middle panel of Fig.~\ref{Spectra1_df}; cross), where a $\sim50$ \% increase in intensity was measured. The intensity at the tip of the arm in the \ion{Si}{IV} $1394$ \AA\ line (top right panel of Fig.~\ref{Spectra1_df}; triangle) at the secondary downflow peak was initially higher than the intensity at the rest wavelength, however, no downflow is detected at this location in later rasters. The different $y$-axes on each plot should be noted by the reader.

\begin{figure*}
\includegraphics[width=0.99\textwidth]{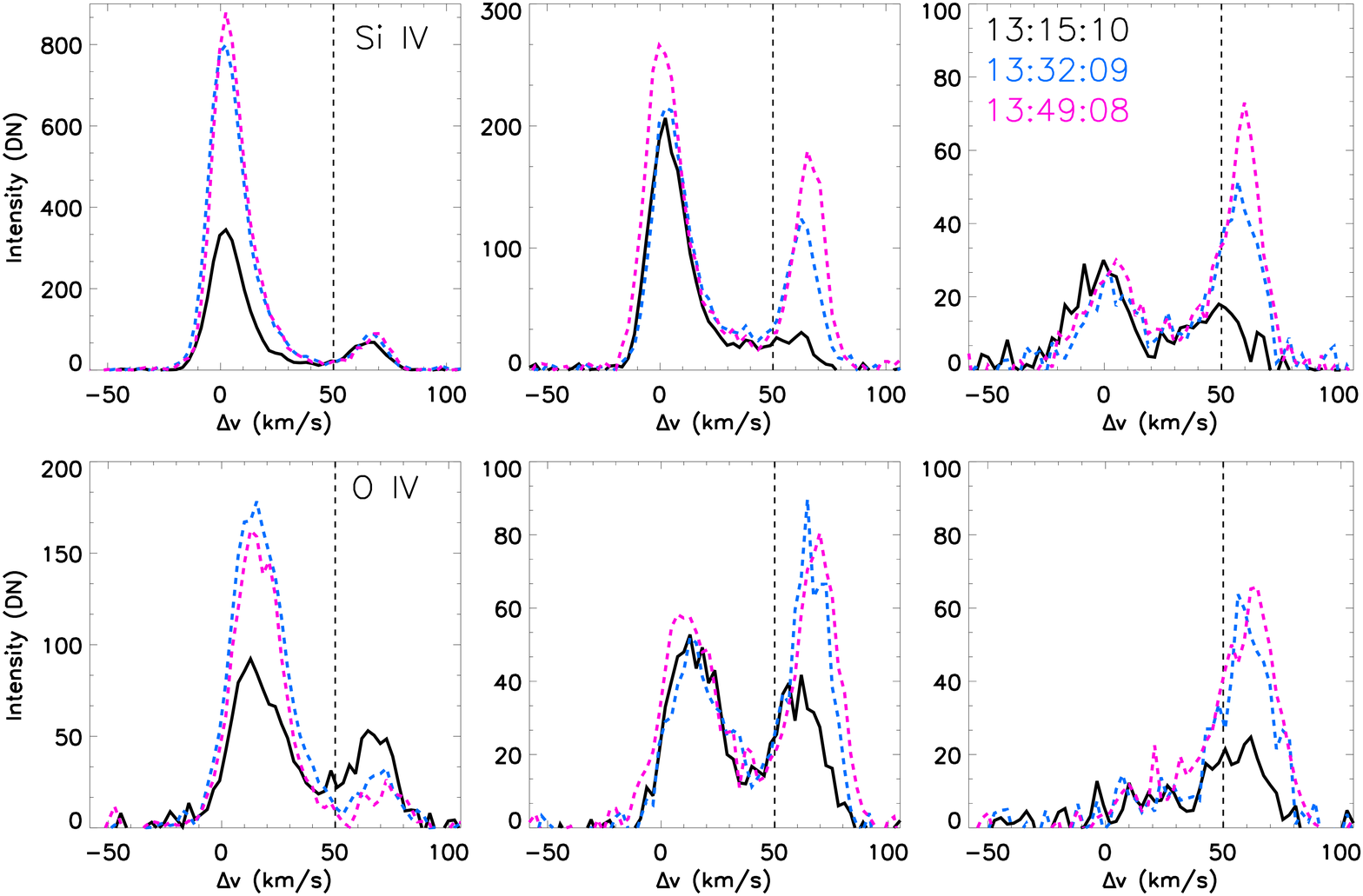}
\caption{Same as Fig.~\ref{Spectra1_df} but for Event (b).}
\label{Spectra2_df}
\end{figure*}

Another interesting result highlighted by Fig.~\ref{Spectra1_df} is that the apparent velocity of the downflow event decreases as one moves away from the apparent foot-point of the feature in the IRIS SJI $1400$ \AA\ data (bottom left to top right for the symbols over-laid on Fig.~\ref{Evolution1_df}). The peak apparent velocity for this feature, over the $50$ km s$^{-1}$ threshold used in this study, was measured at the foot-point of the event (left-hand panels of Fig.~\ref{Spectra1_df}) for both the \ion{Si}{IV} $1394$ \AA\ and \ion{O}{IV} $1401$ \AA\ lines. Around $1$\arcsec\ further along the arm (middle column of Fig.~\ref{Spectra1_df}), however, the apparent velocity of the downflow had dropped below the $50$ km s$^{-1}$ threshold in the \ion{Si}{IV} $1394$ \AA\ line. Another $1$\arcsec\ along the arm and the measured apparent velocity in the \ion{Si}{IV} $1394$ \AA\ line had dropped to around $40$ km s$^{-1}$, well below the cut-off for being considered super-sonic here. Downflows of less than $50$ km s$^{-1}$ were detected all the way to the top of the fingers (right-hand panels of Fig.~\ref{Spectra1_df}) and beyond to the edge of the raster field-of-view (FOV), with the apparent velocity continuing to decrease. We are not able to comment on whether this apparent change in velocity is caused by a real reduction in the velocity or due to projection effects.

\subsubsection{Event (b) - appearance of a transition region downflow}

The second downflow event analysed here, `Event (b)', was located in the centre of the umbra during the rasters sampled on the $21$st May $2015$ and evolved over the final six rasters sampled during this sequence. In Fig.~\ref{Evolution2_df} we plot the equivalent of Fig.~\ref{Evolution1_df} but for Event (b). In the \ion{Si}{IV} $1400$ \AA\ SJI channel (top row) a compact bright region developed at approximately $x_\mathrm{c}$=$405$\arcsec, $y_\mathrm{c}$=$-156$\arcsec\ (co-ordinates from the final column) over these $68$ minutes, with a width of around $1.5$\arcsec\ at the time of the final raster. Several thin fingers (with widths of <$1$\arcsec) were observed to emanate from this bright region in the fourth and fifth columns, each extending towards the bottom right corners of each panel. These fingers are enhanced using a logarithmic scaling in Fig.~\ref{Loops_df}. In the \ion{Si}{IV} $1394$ \AA\ line core (second row), a bright region which increased in size and intensity through time was observed co-spatial to the brightening detected in the SJI data, however, no evidence of the extended fingers was found in this spectral window.  

In the red wing of the \ion{Si}{IV} $1394$ \AA\ line (middle row of Fig.~\ref{Evolution2_df}), the development of the downflow can be observed in the final three columns. Initially (in the first two columns), there was no evidence of any downflows in the spectral window plotted here. After $34$ minutes, however, a region of increased intensity in this velocity window, highlighting the presence of an apparent super-sonic downflow, developed co-spatial to the brightening evident in the \ion{Si}{IV} $1394$ \AA\ line core. After a further $17$ minutes, a large, circular region (with a diameter of around $2$\arcsec) of downflow had appeared extending away from the line core brightening. In the final column, two distinct arms became apparent with each arm forming at the foot-points of the fingers detected in the SJI images. The behaviour in the \ion{O}{IV} $1401$ \AA\ line (fourth and fifth rows for the line core and red wing, respectively) was qualitatively similar to the behaviour observed in the \ion{Si}{IV} $1394$ \AA\ line, but with slightly larger spatial scales (as has been previously noted by \citealt{Nelson20}) and the arms being observed slightly earlier in the fourth column. 

\begin{figure*}
\includegraphics[width=0.99\textwidth]{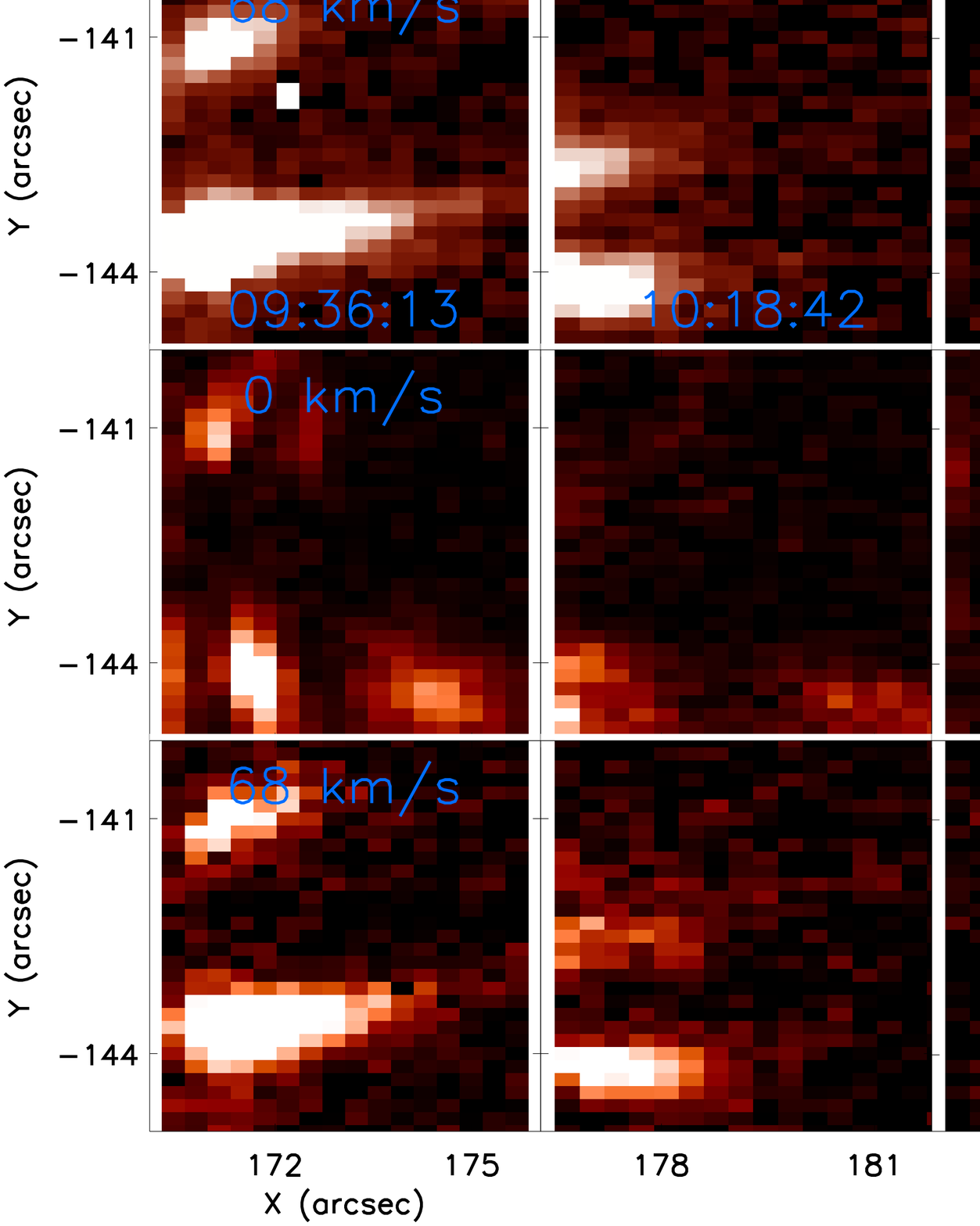}
\caption{Same as Figs.~\ref{Evolution1_df} and \ref{Evolution2_df} but for Event (c) and with the red wing images being sampled at $68$ km s$^{-1}$ for the \ion{Si}{IV} $1394$ \AA\ and \ion{O}{IV} $1401$ \AA\ lines, respectively.}
\label{Evolution4_df}
\end{figure*}

In Fig.~\ref{Spectra2_df} we plot a spectral representation of the evolution of Event (b) at the locations indicated by the symbols on the right-middle panel of Fig.~\ref{Evolution2_df}. As with Event (a), the apparent velocity of this downflow was spatially dependent, with the peak apparent velocity of around $70$ km s$^{-1}$ being measured at the foot-point of the event in the \ion{Si}{IV} $1394$ \AA\ line (top left panel of Fig.~\ref{Spectra2_df}; star). The co-spatial and co-temporal \ion{O}{IV} $1401$ \AA\ apparent velocity was also close to $70$ km s$^{-1}$ (bottom left panel of Fig.~\ref{Spectra2_df}). Half way along the super-sonic downflow (middle panels of Fig.~\ref{Spectra2_df}; cross), the apparent velocity reduced to around $65$ km s$^{-1}$ in both the \ion{Si}{IV} $1394$ \AA\ and \ion{O}{IV} $1401$ \AA\ lines. At the tip of the event, the apparent downflow velocity had reduced again to around $60$ km s$^{-1}$. Again, it is not possible to infer whether this spatial structuring is caused by a real change in the velocity or due to projection effects. The more extended nature of the brightening in the line wing compared to the line core can be observed easily at the tip of Event (b) (right-hand panels of Fig.~\ref{Spectra2_df}; triangle). 

In addition to being spatially dependent, the apparent velocity within Event (b) was also temporally dependent. At the foot-point of the event, the apparent velocity inferred from the \ion{Si}{IV} $1394$ \AA\ line increased from $65$ km s$^{-1}$ to $70$ km s$^{-1}$ over the $34$ minutes plotted in Fig.~\ref{Spectra2_df}. At the tip of Event (b), however, the apparent acceleration rate was higher, with the inferred velocity in the \ion{Si}{IV} $1394$ \AA\ line increasing from around $50$ km s$^{-1}$ to approximately $60$ km s$^{-1}$. The most obvious spectral representation of this apparent acceleration can be found in the middle panel of the \ion{O}{IV} $1401$ \AA\ row of Fig.~\ref{Spectra2_df}. These measured apparent acceleration rates, of around $3$-$5$ m s$^{-2}$, are similar to the rates inferred during the on-set of downflows in the \ion{Si}{IV} $1394$ \AA\ and \ion{O}{IV} $1401$ \AA\ lines in previous research (\citealt{Chitta16, Nelson20}). As with Event (a), no secondary emission peak was detected in the \ion{Mg}{II} or \ion{C}{II} lines.

\begin{figure*}
\includegraphics[width=0.99\textwidth]{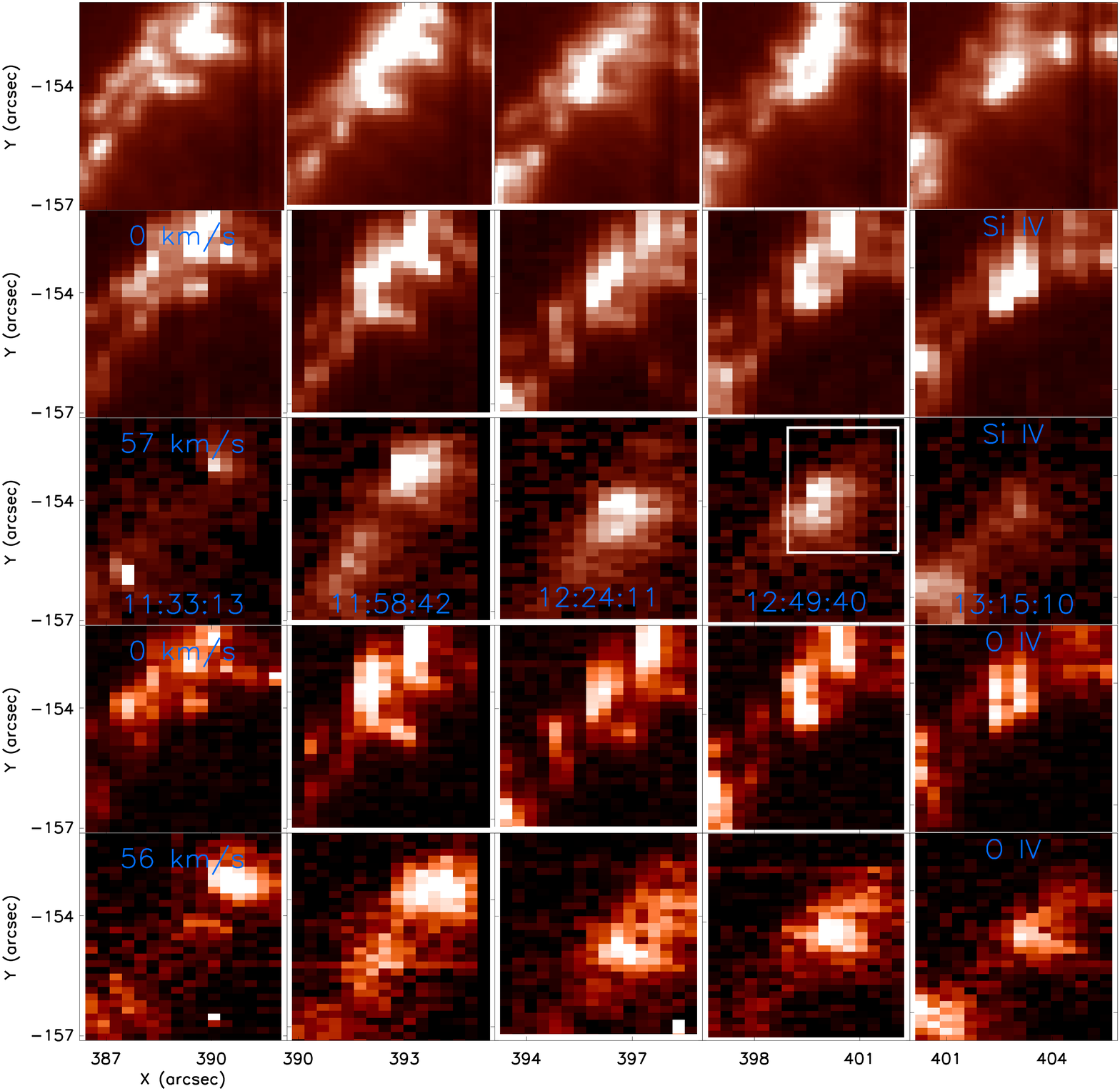}
\caption{Same as Figs.~\ref{Evolution1_df}, \ref{Evolution2_df}, and \ref{Evolution4_df} but for Event (d) and with the red wing images being sampled at $57$ km s$^{-1}$ and $56$ km s$^{-1}$ for the \ion{Si}{IV} $1394$ \AA\ and \ion{O}{IV} $1401$ \AA\ lines, respectively. The white box outlines the FOV plotted in Fig.~\ref{Sub_sonic_df}.}
\label{Evolution3_df}
\end{figure*}

\subsubsection{Event (c) - recurrence of a transition region downflow}

\begin{figure}
\includegraphics[width=0.99\columnwidth,trim={0 1cm 0 0}]{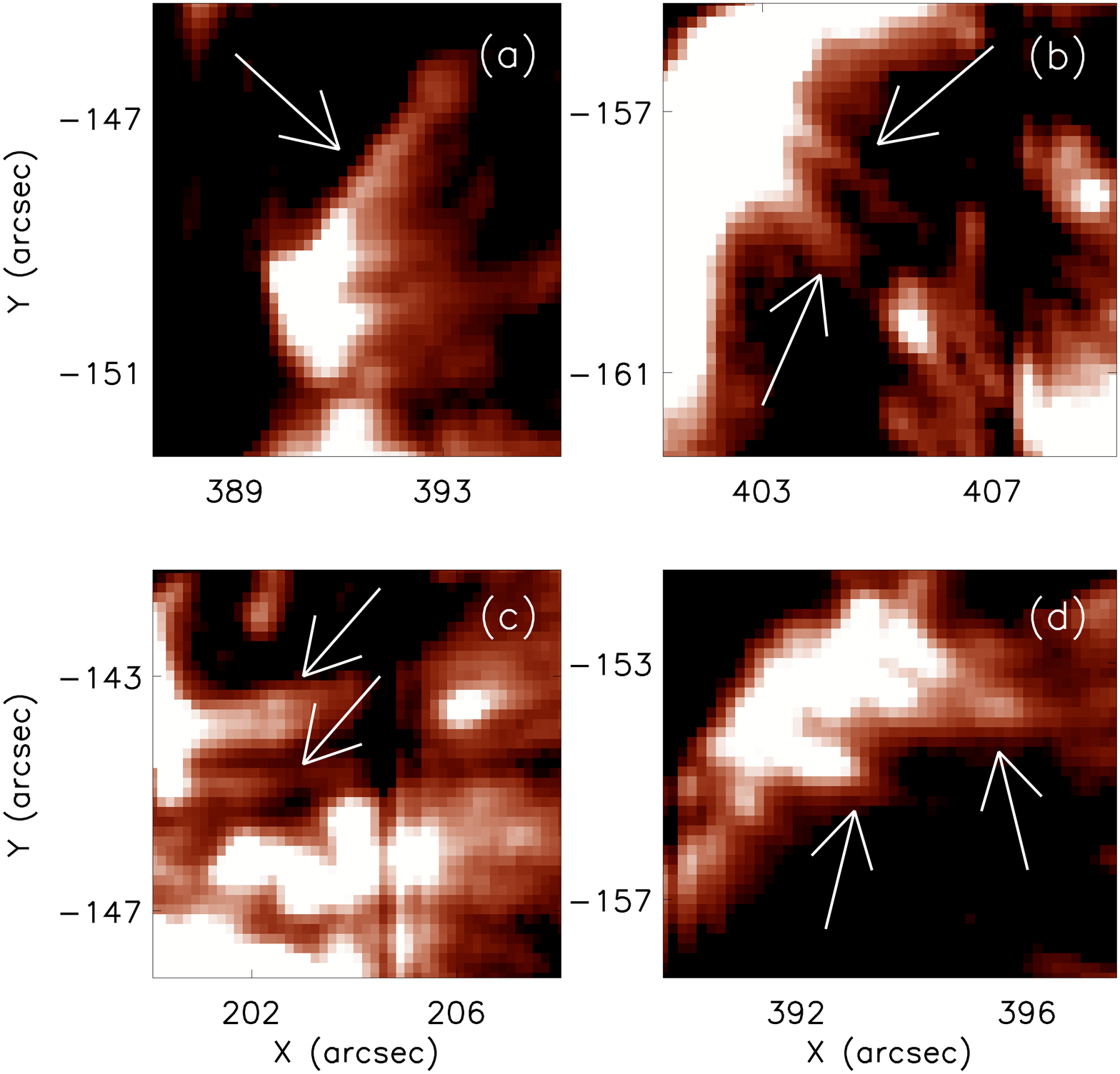}
\caption{Logarithmically scaled images sampled by the IRIS \ion{Si}{IV} $1400$ \AA\ channel for each of the events discussed in this article. The white arrows locate the extended fingers discussed in the text.}
\label{Loops_df}
\end{figure}

Event (c) was the only downflow event identified in the dataset sampled on the $20$th May $2015$ and was the spatially largest and longest-lived downflow studied in this article. This downflow, with a peak apparent velocity of approximately $70$ km s$^{-1}$ in both the \ion{Si}{IV} $1394$ \AA\ and \ion{O}{IV} $1401$ \AA\ lines, was present close to the umbra-penumbra boundary in both the first and final rasters sampled during this observing sequence (separated by more than four hours) and manifested as several long, thin arms (either one, two, or three when it was visible in these data) of downflow with widths of around $1$-$2$\arcsec. In Fig.~\ref{Evolution4_df}, we plot the equivalent of Figs.~\ref{Evolution1_df} and \ref{Evolution2_df} but for Event (c). Initially in the SJI data (top row), an extended brightening reminiscent of Event (a) was present at the north western portion of the plotted FOV and a large region ($3$\arcsec\ diameter) of increased brightening was present in the south western section of the FOV. Over the following ten rasters, the size and intensity of both of these regions decreased (second and third columns) until they were completely absent. Around $30$ minutes later, a new region of increased brightening appeared in the southern portion of the FOV (fourth and final columns). Although extended fingers are not evident in the FOV plotted here, such features can be identified in some SJ images (see Fig.~\ref{Loops_df}). The evolution of Event (c) in the \ion{Si}{IV} $1394$ \AA\ (second row) and \ion{O}{IV} $1401$ \AA\ (fourth row) line cores is similar to that observed in the SJI data.

\begin{figure*}
\includegraphics[width=0.99\textwidth]{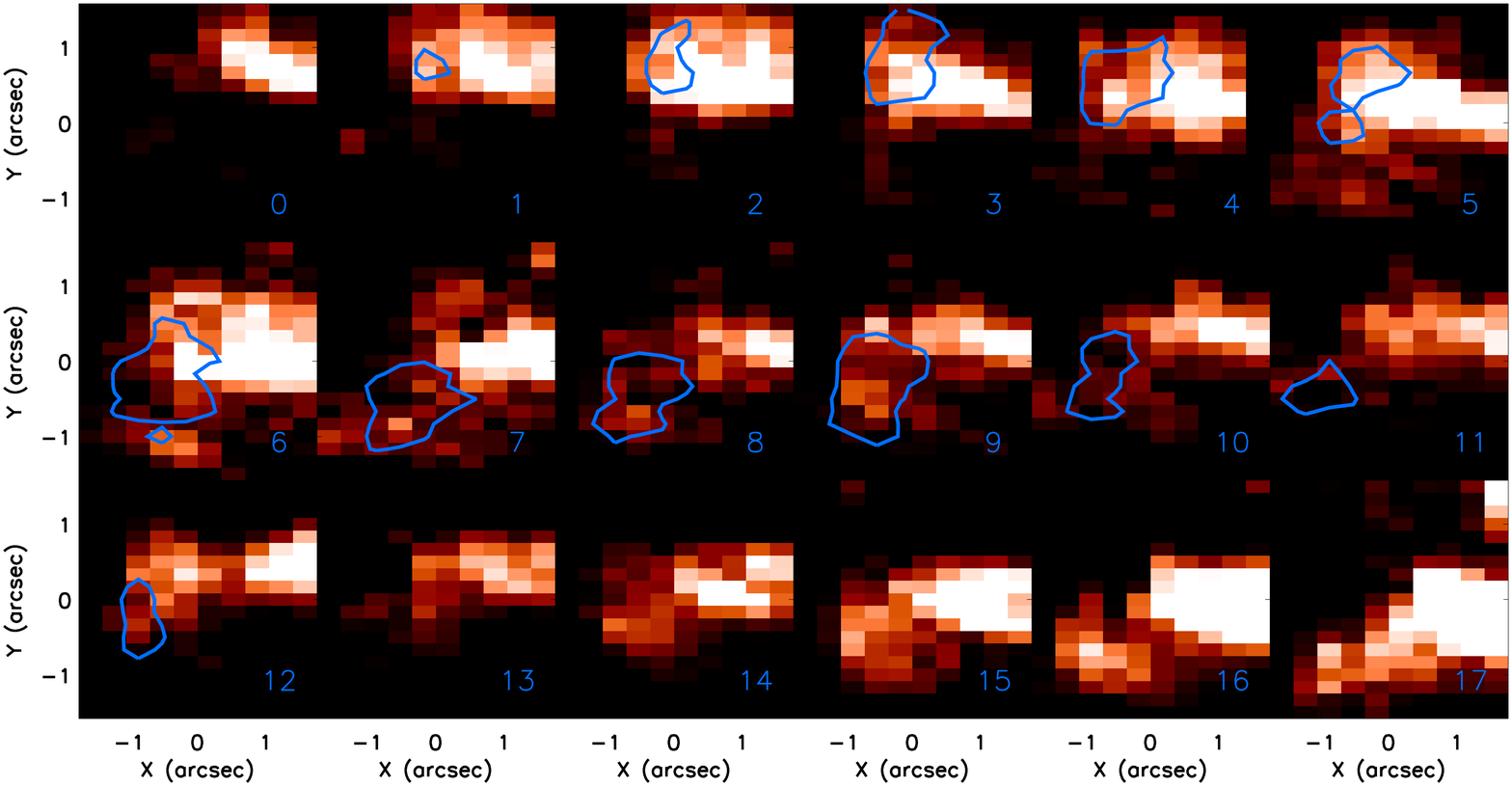}
\caption{The intensity measured at the spectral position corresponding to an apparent velocity of $38$ km s$^{-1}$ into the red wing of the \ion{Si}{IV} $1394$ \AA\ line for each of the $18$ rasters sampled during the $21$st May $2015$ observing sequence. The blue contours over-laid highlight regions where increased intensity (above $40$ DN) was detected at the spectral position corresponding to an apparent velocity of $57$ km s$^{-1}$. The FOV of this plot corresponds to the white box over-laid on Fig.~\ref{Evolution3_df}.}
\label{Sub_sonic_df}
\end{figure*}

The main aspect of Event (c) that we focus on here is the apparent disappearance and then reappearance of the `dual-flow' downflow signatures at this spatial location in the red wing of the \ion{Si}{IV} $1394$ \AA\ (middle row) and \ion{O}{IV} $1401$ \AA\ lines. In the first raster of the time-series collected on the $20$th May $2015$, sampled at around $09$:$27$:$43$ UT, Event (c) was detectable as two distinct regions of  downflow with lengths of around $3$-$5$\arcsec\ and separations of approximately $2$\arcsec. Over the next $90$ minutes these downflows evolve from raster to raster, gradually reducing in size (second column) before they completely disappear from view in the raster beginning at $11$:$01$:$10$ UT (middle column). Over the next $34$ minutes, no evidence of super-sonic downflows is present in these rasters, however, by $11$:$43$:$39$ UT a small region of super-sonic downflow ($<2$\arcsec\ in length) was present (fourth column). This region increased in length until $13$:$00$:$06$ UT (fifth column) before it appeared to split to form two separate arms which were present until the final raster collected at $13$:$34$:$05$ UT.  As with Events (a) and (b), no evidence of this downflow was present in the red wings of the \ion{C}{II} or \ion{Mg}{II} lines. Additionally, no discernable changes occurred in the Hinode/SOT G-band and \ion{Ca}{II} H data during the disappearance and reappearance of the downflow in the transition region. This is perhaps to be expected though due to the wide-band nature of these filters.

\subsubsection{Event (d) - links to sub-sonic downflows}

Whereas Events (a), (b), and (c) could easily be defined spatially, Event (d) was much more complex with its structuring appearing to evolve considerably above the umbra over short time-scales in the rasters sampled on the $21$st May $2015$. In Fig.~\ref{Evolution3_df}, we plot the equivalent of Figs.~\ref{Evolution1_df}, \ref{Evolution2_df}, and \ref{Evolution4_df} but for Event (d). In the \ion{Si}{IV} $1400$ \AA\ SJI channel and \ion{Si}{IV} $1394$ \AA\ line core rows (top and second rows, respectively), a long (around $4$\arcsec\ in length) region of brightness was detected, corresponding to the transition region lightbridge evident in Fig.~\ref{Overview_df}. The structuring of this brightening varied from column to column (separated temporally by around $25$ minutes) with extended fingers being detected to protrude towards the bottom right corner of the FOV in several panels (most notably in the second column). Although the spatial structuring of this event was dynamic, the average intensity and total area of the brightening detected in the \ion{Si}{IV} $1394$ \AA\ line remained relatively consistent through time meaning inferences about links between the line core brightening and any super-sonic downflows present at this location are, unfortunately, impossible at this time. 

Two localised regions of downflow were evident at the bottom left and top right of the FOV in the first column of Fig.~\ref{Evolution3_df}, which was constructed using the second raster sampled in this time-series.  These regions (situated at approximate co-ordinates of [$387$\arcsec, $-156$\arcsec] and [$390$\arcsec, $-153$\arcsec]) were separated by around $4$\arcsec\ and had widths of close to $1$\arcsec. No evidence of these downflows was present in the first raster which was collected around $8$ minutes beforehand. Over the next three rasters, both regions of downflow became more diffuse along the brightening detected in the \ion{Si}{IV} $1394$ \AA\ line core, expanding to diameters of around $2$\arcsec. Whereas the southern region of downflow then faded from view, the northern downflow increased in brightness and was sustained over the course of the next $50$ minutes (third and fourth columns). Finally, the northern downflow began to reduce in size and intensity (final column) eventually fading from view entirely at around $13$:$23$ UT. 

\begin{figure*}
\includegraphics[width=0.99\textwidth]{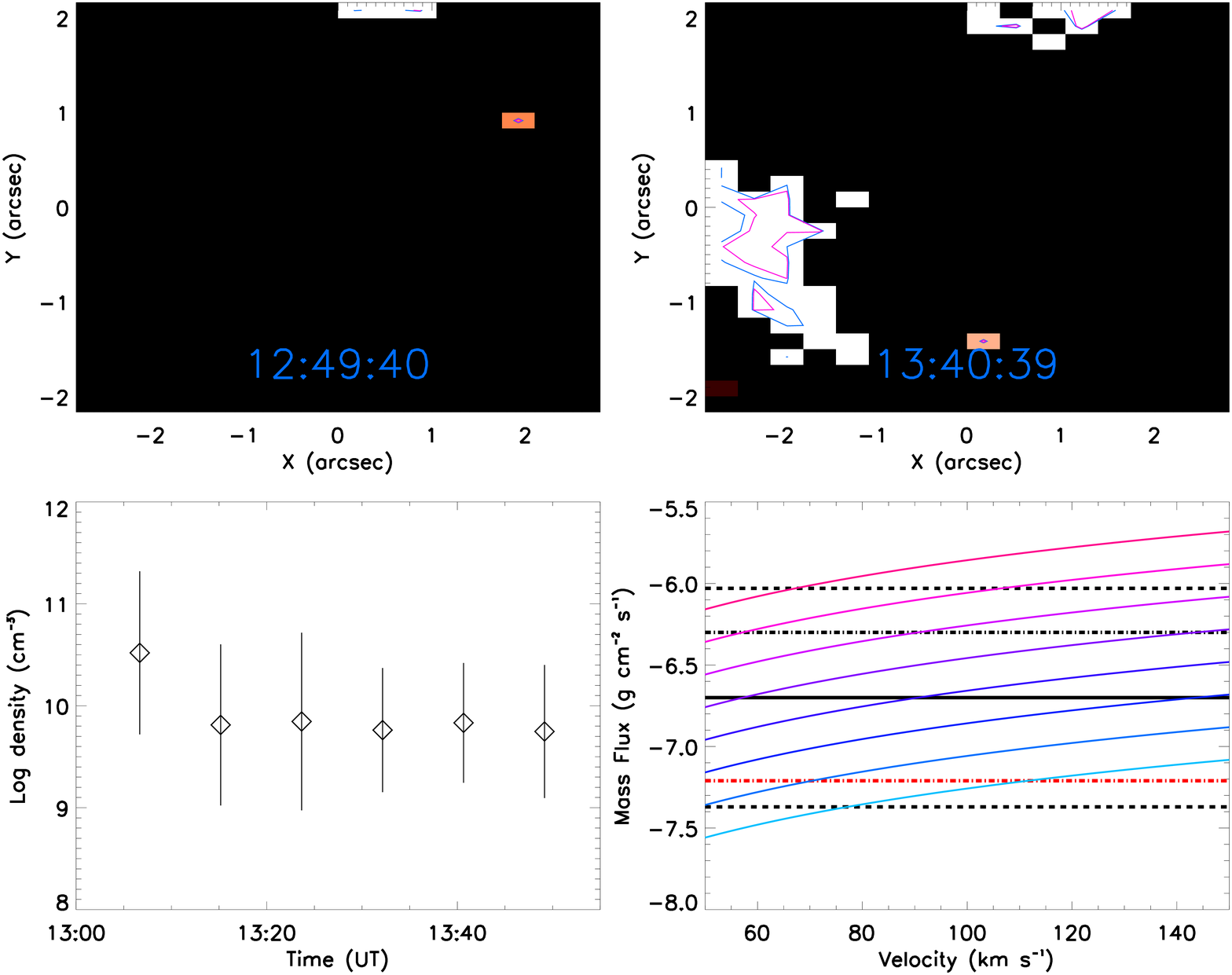}
\caption{(Top row) Maps for Event (b) constructed by summing the intensities within the spectral window [$53$ km s$^{-1}$, $81$ km s$^{-1}$] into the red wing of the \ion{O}{IV} $1401$ \AA\ line $51$ minutes apart. The left-hand panel clearly depicts the lack of downflow at that time. The contours over-laid on these plots outline regions where the logarithmic electron density inferred from the secondary emission peaks in the red wing of the \ion{O}{IV} $1400$ \AA\ and \ion{O}{IV} $1401$ \AA\ lines are above $9.8$ cm$^{-3}$ (blue) and $10.17$ cm$^{-3}$ (pink). (Bottom left panel) The mean logarithmic electron density (diamonds) for each of the rasters when Event (b) was observable. The vertical lines indicate the standard deviations in each raster. (Bottom right panel) Parametric study for the relationship between velocity/density and the mass flux associated with these super-sonic downflows. From aqua to pink (or bottom to top), the input logarithmic electron density transitions from $9.6$ cm$^{-3}$ to $11$ cm$^{-2}$ in jumps of $0.2$ cm$^{-3}$. The dot-dashed black line corresponds to the mass flux reported by \citealt{Straus15}, the solid (dashed) black line correspond to the mean (mean $\pm$ one standard deviation) reported in \citealt{Samanta18}, and the red dot-dashed line corresponds to the mass flux found for Event (b) here.}
\label{Density_df}
\end{figure*}

The behaviour of this downflow in the \ion{O}{IV} $1401$ \AA\ line core (fourth row of Fig.~\ref{Evolution3_df}) was qualitatitively similar to the behaviour described for the \ion{Si}{IV} $1394$ \AA\ line. For the line wing, however, no downflow is detected at the southern region of Event (d) in the first column. Additionally, the initial area coverage of the downflow at the northern region is much larger in the \ion{O}{IV} $1401$ \AA\ line than the \ion{Si}{IV} $1394$ \AA\ line. This result reinforces previous research which has found that the spatial structuring of downflows in the \ion{Si}{IV} $1394$ \AA\ and \ion{O}{IV} $1401$ \AA\ lines is often different (see \citealt{Nelson20}). No signature of this downflow was detected in the red wings of the \ion{Mg}{II} and \ion{C}{II} lines. The on-set of Event (b) can be detected in the bottom left corner of the final \ion{Si}{IV} $1394$ \AA\ red wing panel.

Although Event (d) was identified and studied due to the presence of apparent super-sonic velocities at this location, one of the most interesting aspects of this feature occurred close to the northern downflow event at much lower red-shifts. In Fig.~\ref{Sub_sonic_df}, we plot the intensity at the spectral position corresponding to approximately $38$ km s$^{-1}$ into the red wing of the \ion{Si}{IV} $1394$ \AA\ line for each of the $18$ rasters sampled during this observing sequence within the FOV outlined by the white box over-laid on Fig.~\ref{Evolution3_df}. The downflow present in this velocity window (which displayed the same 'dual velocity' spectral structure as the higher velocity downflows) extended out beyond the edge of the raster FOV, with a similar orientation to the fingers evident in the \ion{Si}{IV} $1394$ \AA\ line core. The lifetime of this sub-sonic component was much larger than the lifetime measured for the higher velocity downflow (blue contours over-laid on each panel of Fig.~\ref{Sub_sonic_df} indicate locations where intensities above $40$ DN were detected at $57$ km s$^{-1}$ into the red wing) and displayed different temporal behaviour, with the largest (both in terms of spatial coverage and secondary peak intensity) downflows at this location being detected in the final three rasters, after Event (d) had disappeared at higher velocities. This sub-sonic downflow was also present in the \ion{O}{IV} $1401$ \AA\ line, with an apparent velocity comparable to that measured in the \ion{Si}{IV} $1394$ \AA\ line.

\subsection{Density estimates and mass flux}

Analysis of the ratio between the \ion{O}{IV} $1400$ \AA\ and \ion{O}{IV} $1401$ \AA\ lines allows important information to be inferred regarding the electron density in the transition region (\citealt{Dwivedi92}). In this study we investigated the evolution of densities within these events using these well-studied lines. Event (b) was selected as an example to plot as it was the largest single event (i.e. only contained one clear region of downflow) studied here. Throughout this analysis, we ignored the potential blend between the \ion{O}{IV} $1401$ \AA\ line and the \ion{S}{I} $1401.51$ \AA\ line, as this is typically weak within sunspots (\citealt{Keenan02}), and focused exclusively on the secondary emission peaks of the \ion{O}{IV} lines in order to investigate the densities within the apparent super-sonic downflow components themselves. Line intensities of the secondary emission peaks were calculated for both \ion{O}{IV} lines by summing the counts in spectral windows covering the velocity range [$53$ km s$^{-1}$, $81$ km s$^{-1}$]. Only pixels which contained a peak in the \ion{O}{IV} $1401$ \AA\ line of between $10$ and $100$ counts (DN) within this range were considered for analysis. The upper limit was introduced to remove any potential influence from spiked pixels. The ratio between these two lines was then calculated and converted to electron densities using values interpolated from those reported in Table~2 of \citet{Young18}.

In the top row of Fig.~\ref{Density_df}, we plot the summed intensity within the spectral window studied for the \ion{O}{IV} $1401$ \AA\ line for two rasters sampled approximately $50$ minutes apart. In the left-hand panel, no increased intensity is detected, however, in the right-hand panel a bright region corresponding to Event (b) is apparent. The diamonds on the bottom left panel of Fig.~\ref{Density_df} plot the evolution of the mean inferred electron density across the spatial extent of Event (b) through time. The inferred electron density, $log(n_\mathrm{e})$, within this feature remained relatively stable during these scans, at a value of approximately $9.8$ cm$^{-3}$. This value is towards the lower end of electron densities previously reported for super-sonic downflows in sunspots (\citealt{Samanta18}).  The vertical lines on the bottom left panel of Fig.~\ref{Density_df} plot one standard deviation as calculated from the inferred electron densities in each raster. The contours over-laid on the top panels of Fig.~\ref{Density_df} outline regions where the logarithmic electron density from the secondary emission peaks is above $9.8$ cm$^{-3}$ (the average density inferred here; blue contours) and $10.17$ cm$^{-3}$ (the average value inferred by \citealt{Samanta18}; pink contours). The logarithmic densities calculated using a similar method for Events (a), (c), and (d) were $9.6$ cm$^{-3}$, $10.2$ cm$^{-3}$, and $9.6$ cm$^{-3}$, respectively.

The mass flux, $F$, associated with Event (b) was calculated using the formula $F$=$n_\mathrm{e}m_\mathrm{p}(n_\mathrm{H}/n_\mathrm{e})v_\mathrm{D}$, where $n_\mathrm{e}$ is the inferred electron density, $m_\mathrm{p}$ is the proton mass, and $v_\mathrm{D}$ is the measured apparent downflow velocity. The term $(n_\mathrm{H}/n_\mathrm{e})$ corresponds to the ratio between the hydrogen and electron densities which is assumed to be $0.83$ in line with previous research (see, for example, \citealt{Straus15, Samanta18}). Inserting the densities and apparent velocities returned for Event (b) into this equation returned a mass flux of $10^{-7.21}$ g cm$^{-2}$ s$^{-1}$ for Event (b), which is seemingly low for super-sonic downflows (\citealt{Samanta18}). Mass fluxes for Events (a), (c), and (d) were calculated as $10^{-7.45}$ g cm$^{-2}$ s$^{-1}$, $10^{-6.81}$ g cm$^{-2}$ s$^{-1}$, and $10^{-7.48}$ g cm$^{-2}$ s$^{-1}$, respectively. As the methods used here contain numerous errors which are difficult to quantify (including both the density and velocity), we conducted a parameter study to further investigate the effects of both velocity and density on the inferred mass flux. In the bottom right panel of Fig.~\ref{Density_df} we plot the relationship between velocity and mass flux for eight different downflow densities (spanning from $log(n_\mathrm{e})$=$9.6$ cm$^{-3}$ [aqua blue] to $log(n_\mathrm{e})$=$11$ cm$^{-3}$ [pink] in jumps of $0.2$ cm$^{-3}$). As expected, the estimated mass flux increases with both velocity and density. The horizontal lines over-laid on this plot indicate the range of values reported in the literature. The solid and dashed black lines indicate the mean and mean $\pm$ one standard deviation inferred by \citet{Samanta18}, respectively. The dot-dashed black line indicates the mass flux in the super-sonic downflow inferred by \citet{Straus15} for one stable super-sonic downflow. The red dot-dashed line indicates the value inferred for Event (b) in this work. It is clear that logarithmic electron densities outside of the range [$9.6$, $10.8$] cm$^{-3}$ predominantly return mass flux estimates for super-sonic downflows outside of typical values (black dashed lines). These values provide some constraints for future modelling efforts.

\subsection{Potential links to coronal loops}

As links between coronal loops and transition region downflows have already been presented in the literature (see, for example, \citealt{Chitta16, Nelson20}), finally we investigated any such relationships for the events studied here. The apparent co-spatial nature of the transition region brightening that spanned across the umbra of the sunspot (and hosted the downflows researched in this article) and the array of fan loops detectable in the SDO/AIA $171$ \AA\ filter to the east of the sunspot was displayed in Fig.~\ref{Overview_df}; however, as the SDO/AIA instrument is not able to resolve individual loops we instead focused on the IRIS SJI channels for the research discussed here. In Fig.~\ref{Loops_df}, we plot an extended FOV, logarithmically scaled, around each super-sonic downflow for the \ion{Si}{IV} $1400$ \AA\ channel. White arrows indicate the fingers in each panel. The widths of each fingers in the \ion{Si}{IV} $1394$ \AA\ channel was less than $1$\arcsec\ (similar to the inferred widths of coronal loops in the highest resolution coronal imaging currently available; \citealt{Williams20}) and the lengths were between $3$-$5$\arcsec. The lifetimes of these fingers ranged from a few minutes to more than half an hour. After careful analysis, no signature of these fingers was detected in the \ion{Mg}{II} $2796$ \AA\ filter.

Examination of the SDO/AIA coronal lines revealed that the arms detected in the IRIS SJI $1400$ \AA\ data extended out in the same direction as the coronal fan loops evident in Fig.~\ref{Overview_df}. Unfortunately, as these extended fingers have widths below the spatial resolution of the SDO/AIA instrument and cannot be detected more than approximately $5$\arcsec\ away from the super-sonic components of the downflows, it is not possible to conclusively link them to coronal loops, neither individually or in general, in this work. Analysis of the G-band and \ion{Ca}{II} H imaging data sampled by Hinode/SOT co-temporally to Event (c) provided no evidence to link these downflows to other features in either the photosphere (e.g., umbral dots) or chromosphere (e.g., penumbral filaments). This may be expected, however, due to the wide-band nature of these filters. One example of future research would involve investigating whether downflows could be detected significant distances ($10$s of arcseconds) away from the sunspots in the \ion{Si}{IV} $1394$ \AA\ line in certain circumstances (i.e., specific magnetic topologies at specific observing angles). Such work would allow us to better investigate the structures these downflows form within. Another promising direction of future research may involve the combination of IRIS data with coronal imaging sampled by the High-Resolution Coronal Imager (Hi-C; \citealt{Kobayashi14}). Hi-C is capable of observing the solar corona at resolutions close to the IRIS SJI meaning discovering correspondences between fine-scale structures in the transition region and corona could be possible.

\section{Discussion and Conclusions}
	\label{Conclusions}

In this article, we studied four distinct downflows with peak secondary emission components at Doppler velocities above $50$ km s$^{-1}$ in both the \ion{Si}{IV} $1394$ \AA\ and \ion{O}{IV} $1401$ \AA\ lines. The apparent super-sonic components of each of these downflows had widths of around $2$-$3$\arcsec, but displayed some evidence of fine-scale structuring including various arms at some points during their lifetimes (see Figs.~\ref{Evolution1_df}, \ref{Evolution2_df}, \ref{Evolution4_df}, and \ref{Evolution3_df}). All four events appeared to form exclusively at the foot-points of multiple thin ($<1$\arcsec) `fingers' extending around $3$-$5$\arcsec\ through the \ion{Si}{IV} $1400$ \AA\ SJI channel (Fig.~\ref{Loops_df}) with the same orientation as a group of coronal fan loops observed in the SDO/AIA $171$ \AA\ filter. This potential alignment would agree with results presented in the previous literature (see, for example, \citealt{Chitta16, Nelson20}) and could support the hypothesis that flows within coronal loops are responsible for high velocity transition region downflows in the transition regions of sunspots (as discussed by, for example, \citealt{Kleint14, Ishikawa20}). Alignment between IRIS and Hi-C data in the future would allow further inferences to be made about the relationship between downflows and coronal loops in general.

The high velocity components of each of these downflows were detectable in the \ion{Si}{IV} $1394$ \AA\ red wing for more than $1000$ s, however, only Event (d) was (potentially) observed during its entire existence with a lifetime of more than $4000$ s. Notably, downflows were evident at the same location intermittently for over $14000$ s within Event (c). For Event (a) we only observed the high velocity downflow as it faded from view in the first three to four rasters, whereas for Event (b) we only observed its appearance phase in the final six rasters. Event (c) initially faded from view before recurring at the same location indicating that the driver of these downflows may be repetitive in nature. The peak apparent velocities of these downflows were between $55$ km s$^{-1}$ and $70$ km s$^{-1}$ (see Fig.~\ref{Spectra1_df} and Fig.~\ref{Spectra2_df}) with no disparity in apparent velocities measured from the \ion{Si}{IV} $1394$ \AA\ and \ion{O}{IV} $1401$ \AA\ lines (agreeing with the results of \citealt{Samanta18, Nelson20}). The peak apparent velocities for these events are relatively low compared to previous examples of super-sonic downflows detected using these lines (see, for example, \citealt{Straus15, Chitta16, Samanta18, Nelson20}).  Some acceleration was detected during the appearance phase of Event (b) with similar value to accelerations previously reported in the literature (see, for example, \citealt{Chitta16, Nelson20}), however, no clear evidence of deceleration was present during the disappearance of Event (a). Interestingly, a large region of sub-sonic downflow, with an apparent velocity of $38$ km s$^{-1}$, was detected seemingly above Event (d) after its apparent super-sonic component had faded from view (see Fig.~\ref{Sub_sonic_df}) indicating that the higher velocity super-sonic downflows could be linked to larger-scale flow structures in the upper solar atmosphere above sunspots.

In order to further investigate these downflows, we analysed the density and mass flux associated with these events through time (Fig.~\ref{Density_df}) using the ratio between the \ion{O}{IV} $1400$ \AA\ and \ion{O}{IV} $1401$ \AA\ lines (for further details about this ratio see, for example, \citealt{Dwivedi92, Young18}). Specifically, we calculated this ratio using the total intensity of the secondary emission peaks in the red wings of these lines in order to analyse the high velocity downflows themselves. The inferred logarithmic densities, $log(n_\mathrm{e})$, lay within the range around $9.6$-$10.2$ cm $^{-3}$ which returned mass flux estimates of between $10^{-6.81}$-$10^{-7.21}$ g cm$^{-2}$ s$^{-1}$. Both of these ranges are on the low side when compared to the mean found for super-sonic downflows in previous research (\citealt{Samanta18}). After conducting a parametric study, we propose that future efforts model these super-sonic downflows should begin using an electron density in the range $log(n_\mathrm{e})$=[$9.6$, $10.8$] cm$^{-3}$.

Finally, we briefly discuss the implications these results have on the formation mechanisms of downflows with apparent super-sonic velocities in the transition region. Previous authors have suggested that mass draining from a standard coronal loop, potentially due to siphon (\citealt{Straus15}) or condensation (\citealt{Antolin15}) flows, will occur over time-scales of $<1000$ s (\citealt{Straus15, Chitta16}) meaning such a mechanism cannot sustain super-sonic downflows (given their lifetimes are longer than this). However, as the downflows studied here each appear to display several extended fingers and Event (c) indicates that super-sonic downflows can recur at the same spatial location, it may be possible that each super-sonic downflow is linked to multiple elongated coronal rain events. Measurements of the widths of coronal rain events which have timescales, densities, mass fluxes comparable to the downflows studied here (see, for example, \citealt{Antolin15}) match the widths of the arms of super-sonic reported here. Overall, our results do not refute the hypothesis that siphon or condensation flows could be the drivers of these high velocity downflows, but a larger sample of IRIS observations will need to be studied to confirm or deny this.

\section{Summary}
	\label{Summary}

Four downflows in the transition region above a sunspot were analysed in both spectral and imaging data sampled by the IRIS satellite in this article. The high velocity components of each downflow appeared to form in localised regions (widths of a few arcseconds) at the foot-points of extended ($3$-$5$\arcsec\ in length) fingers observed in \ion{Si}{IV} $1400$ \AA\ SJI data. These fingers could be linked to coronal loops (potentially hosting elongated coronal rain), however, further research would be required to confirm or deny this. Additionally, the density range ($log(n_\mathrm{e})$=$9.6$ cm$^{-3}$-$10.2$ cm$^{-3}$) and mass flux range ($10^{-6.81}$-$10^{-7.48}$ g cm$^{-2}$ s$^{-1}$) of these events were within the range presented by \citet{Samanta18} in their statistical analysis.
\begin{itemize}
\item{Event (a) was present in the first raster at the foot-point of a brightening in \ion{Si}{IV} $1400$ \AA\ SJI data. The magnitude of the secondary emission peaks in the red wings of both the \ion{Si}{IV} $1394$ \AA\ and \ion{O}{IV} $1401$ \AA\ lines reduced over the course of the next $25$ minutes until they had fully disappeared without any identifiable change in the apparent velocity.}
\item{Event (b) appeared over the course of the final $6$ rasters (around $44$ minutes), expanding from a point-like event to a region with a width of $2$\arcsec, again at the foot-point of an extended brightening in \ion{Si}{IV} $1400$ \AA\ SJI data. This event appeared to increase in velocity through time with an acceleration consistent with those previously reported in the literature (\citealt{Chitta16, Nelson20}). }
\item{Event (c) was the only dual-peaked transition region downflow structure identified in the dataset sampled on the $20$th May $2015$. This feature initially manifested as multiple thin (diameters of around $1$-$2$\arcsec) fingers of downflow which had lengths of around $3$-$5$\arcsec. The downflow at this location reduced in size over the first $90$ minutes of observations before eventually disappearing and then reoccurring around $30$ minutes later, after which it was present until the end of the observations.}
\item{Event (d) was evident in the wings of the \ion{Si}{IV} $1394$ \AA\ and \ion{O}{IV} $1401$ \AA\ lines for more than one hour. During this time, the downflow evolved from two point-like brightenings to a circular region with a width of around $2$\arcsec\ before fading from view entirely. A large region of sub-sonic downflow was observed to develop in the extended fingers linked to Event (c) after the super-sonic component had faded.}
\end{itemize}

\begin{acknowledgements}
We thank the Science and Technology Facilities Council (STFC) for the support received to conduct this research through grant number: ST/P000304/1. SKP is grateful to the FWO Vlaanderen for a senior postdoctoral fellowship. IRIS is a NASA small explorer mission developed and operated by LMSAL with mission operations executed at NASA Ames Research Center and major contributions to downlink communications funded by ESA and the Norwegian Space Centre. SDO/HMI and SDO/AIA data are courtesy of NASA/SDO and the HMI and AIA science teams. Hinode is a Japanese mission developed and launched by ISAS/JAXA, collaborating with NAOJ as a domestic partner, NASA and STFC (UK) as international partners. Scientific operation of the Hinode mission is conducted by the Hinode science team organized at ISAS/JAXA. This team mainly consists of scientists from institutes in the partner countries. Support for the post-launch operation is provided by JAXA and NAOJ(Japan), STFC (U.K.), NASA, ESA, and NSC (Norway).

\end{acknowledgements}

\bibliographystyle{aa}
\bibliography{Transition_Region_downflows}

\end{document}